\documentclass[pre,aps,superscriptaddress,nofootinbib]{revtex4}

\usepackage{epsfig}
\usepackage{amsfonts}
\usepackage{latexsym}
\usepackage{amsmath}
\usepackage{amssymb}
\usepackage{slashed}
\usepackage{longtable}

\usepackage{graphicx}
\usepackage{hyperref}
\usepackage{pstool}

\numberwithin{equation}{section}

 \def\p{\partial}

\newcommand{\bea}{\begin{eqnarray}}
\newcommand{\eea}{\end{eqnarray}}
\newcommand{\be}{\begin{equation}}
\newcommand{\ee}{\end{equation}}
\newcommand{\ba}{\begin{align}}
\newcommand{\ea}{\end{align}}

\newcommand{\der}[1]{\frac{d}{d #1}}

\newcommand{\tr}{\mbox{tr}}

\newcommand{\bphi}{\bar{\phi}}
\newcommand{\btheta}{\bar{\theta}}
\newcommand{\bc}{\bar{c}}

\newcommand{\dd}[1]{\mathbf{{d}#1}}
\newcommand{\dx}{\mathbf{dx}}

\newcommand{\D}[1]{\left[\mathcal{D}#1\right]}

\def\tr{{\rm Tr}}

\def\Or[#1]{{\text{O}}\left({#1}\right)}
\def\dotl[#1,#2]{\left\langle #1, #2 \right\rangle}
\def\dotlb[#1,#2]{[ #1, #2 ]}
\def\dotp[#1,#2]{(#1) \cdot (#2)}
\def\aff[#1,#2]{\hat{#1}(#2)}
\def\n4sym{{\cal N}=4 SYM}
\def\>{\rangle}
\def\<{\langle}
\def\weight[#1,#2,#3]{\{(#1),#2,#3\}}
\def\ads[#1]{$\text{AdS}_{#1}$}

\newcommand{\cor}[1]{\left\langle #1 \right\rangle}

\newcommand{\cb}{\bar{c}}

\newcommand{\Lb}{\bar{L}}

\newcommand{\dv}[1]{\frac{\delta}{\delta #1}}
\newcommand{\Hb}{\bar{H}}

\numberwithin{equation}{section}
\begin{document}

\preprint{}

\title{Symmetries in the path integral formulation of the Langevin dynamics}

\author{Piotr Sur\'owka}
\author{Piotr Witkowski} \affiliation{Max-Planck-Institut  f\"ur Physik komplexer Systeme, N\"othnitzer Str. 38, 01187 Dresden, Germany}

\begin{abstract}
We study dissipative Langevin dynamics in the path integral formulation using the Martin-Siggia-Rose formalism. The effective action is supersymmetric and we identify the supercharges. In addition we study the transformations generated by superderivatives, which were recently included in the cohomological structure emerging in the dissipative systems. We find that these transformations do not generate Ward identities, which are explicitly broken, however, they lead to universal  sum-rule type identities, which we derive from Schwinger-Dyson equations. We confirm that the above identities hold in an explicit example of the Ornstein-Uhlenbeck process.
\end{abstract}

\maketitle
\section{Introduction}

Low energy effective models are at the core of understanding various physical phenomena. In such a limit only a small number of degrees of freedom is relevant and the ignorance about the microscopic details is parameterized by some effective variables. One manifestation of microscopic dynamics comes through the thermal fluctuations arising from the external heat bath. A way to model fluctuating variables is to introduce stochastic terms in the evolution governed by a differential equation. In the modern language stochastic equations can be formulated in terms of path integrals. This originated from the stochastic quantization procedure \cite{DAMGAARD1987227}, which formulates a Euclidean field theory as the equilibrium limit of a statistical system coupled to thermal fluctuations. A field-theoretic formulation of stochastic dynamics allows one to employ powerful symmetry techniques to derive, e.g., statistical work relations. 
Perhaps the most fundamental stochastic differential equation is the Langevin equation related to diffusion processes. Its properties can be used to understand more complicated stochastic evolution. In this work we will study the over-damped Langevin dynamics with noise that belongs to a subclass of potential or gradient systems \cite{CrossHohenberg}. It is well known that Langevin dynamics can be formulated in terms of path integrals using Martin-Siggia-Rose (MSR) construction. Therefore we can view the Langevin dynamics as a toy model that illustrates more general features of stochastic systems. In fact, Burgers equation is an example of Langevin dynamics, that belongs to a more general nongradient case. The effective action of over-damped, potential Langevin equation, constructed from MSR, possesses a number of symmetries \cite{ParisiSourlas,Chaturvedi1984,Orland,zinn2002quantum,Ovchinnikov2016}. A peculiar feature of these symmetries is that they mix physical and the ghost fields present in the theory. Therefore, it is usually stated that the effective action for Langevin dynamics is supersymmetric. It can be linked to the microscopic Schwinger-Keldysh field theory which gives rise to Langevin dynamics in the classical limit \cite{Kamenev}. Exploring Schwinger-Keldysh approach various microscopic constructions were proposed to elucidate these supersymmetries \cite{Haehl:2015foa,Haehl:2016pec,Haehl:2016uah,Glorioso:2016gsa,Glorioso:2017fpd,Jensen:2017kzi,Haehl:2017zac}. Although similar in spirit they have some differences. For example, an inherent feature of \cite{Haehl:2015foa,Haehl:2016pec,Haehl:2016uah,Haehl:2017zac} is the existence of a dynamical gauge field that corresponds to thermal translations. In turn, the number of supersymmetry generators is enlarged. We use this as a motivation to study the properties of these transformations in a conventional set-up of Langevin dynamics with a fixed gauge field.  We find that some symmetries are broken; however, the transformations still generate universal identities, from which some new equilibrium relations can be deduced. Those identities assume a form of a sum-rule for n-point correlation functions. We exemplify our considerations using explicit computations in the Ornstein-Uhlenbeck process \cite{OrnsteinUhlenbeck}.

\section{Particle on a Schwinger-Keldysh contour}

The Schwinger-Keldysh formalism was developed to calculate the nonequilibrium correlation functions. In equilibrium, to calculate correlation functions we use conventional perturbation theory. However, in nonequilibrium we do not have the usual control over the final state. The time-ordered correlation function reads
\be
i G(\mathbf{x},t;\mathbf{x}',t')= <\Phi (\infty) |_F T[\mathcal{S}(\infty,-\infty)\phi(\mathbf{x},t) \phi ^\dagger(\mathbf{x}',t')] |\Phi (-\infty)>_I,
\ee 
where $\mathcal{S}(\infty,-\infty)$ is the $S-$matrix. Out-of-equilibrium the assumption that the final state differs from initial state only by a phase is broken. A method used to avoid dealing with the quantum state at infinity is to evolve back to the initial state,
\be
|\Phi(-\infty)>_I=S(-\infty,\infty) |\Phi(\infty)>_F,
\ee
and introduce a two-branch contour together with a contour ordering $T_c$. In fact we do not even need to consider infinite past if we know the density matrix at some finite time $t_0$. Then we can evolve our system up to some finite time and back to $t_0$. The evolution contour is now closed in time. Finally, we can include finite temperature effects by adding the imaginary time branch, which will implement the thermal boundary conditions.  We can use the contour to define an effective functional that will generate the relevant correlation functions,
\be \label{Eq:Z}
Z[H_1,H_2]=\int [d\phi _1][d\phi_2]\exp \left[ i (S[\phi_1,H_1]-S[\phi_2,H_2])  \right] \exp[iS_{IF}(\phi_1,\phi_2)],
\ee
where $\phi_i$'s are defined on the upper and lower branches of the contour and $S_{IF}$ is an interaction between different copies of the system. Differentiating with respect to the sources gives a matrix of Green's functions. We see from \eqref{Eq:Z} that the Schwinger-Keldysh construction doubles the degrees of freedom and creates certain redundancies in the description, which can be understood as a gauge symmetry \cite{Haehl:2016pec}. This means that we can make redefinitions of fields and the physical observables remain unchanged. One possible choice of such a redefinition is the Keldysh rotation,
\be \label{Eq:split}
\phi _r =\frac{1}{2}(\phi_1+\phi_2),\qquad \phi_a =\phi_1-\phi_2,
\ee
\be
H_r=\frac{1}{2}(H_1+H_2), \qquad H_a=H_1-H_2.
\ee
$r$-type operators are conjugate to $a$-type sources and vice versa. A consequence of such redefinitions is the existence of symmetry charges  that act on functionals. This symmetry enforces a constraint on the correlation functions. For example, if we align sources, then the cyclicity of the trace and the unitarity imply that the partition function is independent of $H$. Therefore, if $a$-type sources are set to zero, then the variations with respect to $r$ sources must vanish and the partition function becomes topological.  As a result, the Schwinger-Keldysh partition function is invariant under one \cite{Jensen:2017kzi} or two topological BRST charges \cite{Haehl:2015foa,Haehl:2016pec,Haehl:2016uah,Haehl:2017zac}, depending on the approach, which enforce the above constraint. Apart from this topological symmetry the partition function has an additional symmetry if the initial state is thermal, which corresponds to the time evolution in imaginary time,
\be \label{Eq:KMS}
Z[H_1(t_1),H_2(t_2)]=\tr \left( U_2^\dag[H_2(t_2)]e^{\beta \mathcal{H}}U_1[ H_1(t_1-i\beta)]\right).
\ee
From \eqref{Eq:KMS} we can obtain the Kubo-Martin-Schwinger (KMS) condition for the thermal correlators. This symmetry is again generated by one or two KMS charges.  It was noted that the four symmetry generators form an algebra, which has been previously encountered in
the topological field theory literature and goes by the name of the extended $\mathcal{N}_T = 2$ equivariant cohomology algebra \cite{BIRMINGHAM1991129}. Finally, the KMS symmetry is combined with CPT invariance, which is spontaneously broken to obtain dissipative effects.

Let us see how we can formulate a particle dynamics on the Schwinger-Keldysh contour. We start with the action
\be
S[\varphi]=\int dt \left[ \frac{1}{2} \dot \varphi ^2 -V(\varphi) \right],
\ee
and we split the field into two components $\phi_1$ and $\phi_2$ residing on a two-branch contour according to \eqref{Eq:split}. In terms of these new fields, the action takes the form
\be
S[\phi _r, \phi _a]=-\int dt \left[ \phi _a \ddot{\phi_r}- V(\phi _r+2 \phi_a)+V(\phi _r-2 \phi_a)\right],
\ee
where we performed an integration by parts. If we assume that the fluctuations of the $a$ component are small, we can expand the potential terms to get
\be \label{Eq:doubledeffective}
S[\phi _r, \phi _a]=-\int dt  \left[  \phi _a \left(\ddot{\phi_r}+ \frac{\p V(\phi_r)}{\p \phi_r}\right)\right].
\ee
We notice that we can perform the integration over $ \phi _a$ in the partition function,
\be
Z= \mathcal{N} \int [d\phi _r]  \delta \left( \ddot{\phi_r}-\frac{\p V(\phi_r)}{\p \phi_r}\right),
\ee
 which gives the equation of motion for the $r$ field,
\be
\ddot{\phi_r}=-\frac{\p V(\phi_r)}{\p \phi_r}.\label{eq:classical}
\ee
As we will see later this form of the action resembles the effective action obtained for Langevin dynamics without noise.  To obtain the noise contribution one has to carefully take into account quantum fluctuations and take $\hbar\rightarrow 0$ limit \cite{Kamenev}.  Therefore, we can view Langevin equation as coming from the Schwinger-Keldysh construction and we expect it to be invariant precisely under the $\mathcal{N}_T = 2$ symmetry, which we will identify as the Parisi-Sourlas supersymmetry \cite{zinn2002quantum}. 

\section{Stochastic differential equations, MSR formalism and supersymmetry}
In our analysis so far we completely ignored the effects of dissipation and fluctuations. Having in mind the Langevin dynamics we want to include these effects in the effective action. It turns out one can do that using the Martin-Siggia-Rose (MSR) prescription.  In essence one starts with a stochastic differential equation (SDE) with noise
\begin{equation}
\label{eq:SPDE}
E[\phi(x)] = \nu(x),
\end{equation}
where $E(\phi)$ is some differential operator and $\nu(x)$ is a random variable. One must carefully define what does the whole expression mean which is usually done by means of It\^o or Stratanovich calculus in a mathematically consistent way. Assuming this we want to calculate the correlation functions for a stochastic process. An efficient tool to achieve this is to construct a partition function and differentiate with respect to sources. To do that in MSR formalism, one starts with the following identity:
\be\label{Eq:partition}
Z[\nu]=\int [dE] \delta (E(\phi) - \nu)=\int [d\phi] \mathcal{J}(\phi) \delta (E(\phi) - \nu).
\ee
$\mathcal{J}(\phi) =\det \frac{\delta E}{\delta \phi} $ is the Jacobian. In this framework, $\phi$ is not a real function of $x$ but rather a random variable itself.\\
We will assume that the noise fulfils 
\begin{equation}
<\nu(x)\nu(x')> = \frac{2 \Gamma}{\beta}\delta(x-x'),
\label{eq:noiseAss}
\end{equation}
i.e. the white noise, with a Gaussian distribution,
\be
\int \D{\nu} \nu(x)\nu(x')\exp \left(-\frac{\beta}{4 \Gamma} \nu^2 \right) =\frac{2 \Gamma}{\beta}\delta(x-x').
\ee
In the next steps we introduce an auxiliary field $\bphi$ that will give us the delta function in \eqref{Eq:partition} and integrate over the noise. The partition function is
\begin{equation}
Z[H, \Hb, L, \Lb]=\int \D{c}\D{\cb}\D{\bphi}\D{\phi} \exp\left(\int\dx~\Sigma(\phi, \bphi, c, \cb) + \Hb\phi + H\bphi + L\cb + c\Lb\right), \label{eq:PartFu}
\end{equation}
where we expressed the Jacobian as an integral over ghost fields and introduced sources for every field. The effective action $\Sigma$ is given by
\begin{equation}
\Sigma(\phi, \bphi, c, \cb)=-\frac{\Gamma}{\beta} \bphi^2 - i\bphi E(\phi) +c\frac{\delta E}{\delta \phi} \cb.
\label{eq: MSRAction}
\end{equation}
It has three auxiliary fields - one real and two Grassmannian. In addition to that, we see that the form of this action resembles equation \eqref{Eq:doubledeffective} upon identification $\phi_a \rightarrow \bphi$. This strongly suggests that we can interpret stochastic dynamics as emerging from microscopic Schwinger-Keldysh construction. 
Up to this point our considerations are completely general. Now we will restrict our attention to the Langevin dynamics.  If equation \eqref{eq:SPDE} has the form
\begin{equation}
\label{eq:Langeven}
E(\phi) = \partial_t \phi +  \Gamma \dv{\phi} U(\phi),
\end{equation}
the SDE is an over-damped, purely dissipative Langevin equation. This type of equation is valid for one-dimensional domains [i.e., $\phi(t)$ being a position of some particle at time $t$] as well as for fields on multidimensional domains (the potential $U$ can depend also on spatial derivatives of $\phi$, see section 2.1  of \citep{Orland} for an example). We will adopt a notation where $x$ in integrals and arguments denotes all the variables, among which there always is time $t$. The latter variable will often be mentioned separately, and denoted by $t$. This kind of equation is a valid approximation when the inertia of a particle is negligible in comparison to the linear damping force. One physical realization of this dynamics describes the evolution of the order parameter of a second order phase transition in axial ferromagnets.  

The connection between Langevin dynamics and Schwinger-Keldysh field theories suggests that the effective action possesses an underlying $\mathcal{N}_T=2$ symmetry structure \cite{Haehl:2015foa}, which will lead to identities between various correlation functions. This fact was noted a long time ago in the context of dimensional reduction which was later used to unearth various properties and methods to study Langevin dynamics. Explicitly, as shown in \cite{Orland}, the action is invariant under the following transformation:
\begin{equation}
\begin{array}{l l}
Q: & \delta \phi = -\cb \epsilon,~\delta c = i\bar{\phi}\epsilon{},~\text{other variations vanishing}\\
\bar{Q}: & \delta \phi = c \epsilon,~\delta \bc =(i \bphi -\frac{\beta}{\Gamma} \dot{\phi})\epsilon,~\delta \bphi=-i\frac{\beta}{\Gamma} \dot{c} \epsilon,~\text{other variations vanishing}.
\end{array}
\label{eq:SymmetryAlgebra}
\end{equation}
Apart from them, we can define two other operators that complete the algebra, following \cite{Haehl:2015foa}
\begin{equation}
\begin{array}{l l}
D: & \delta \phi = c \epsilon,~\delta \bc = i\bar{\phi}\epsilon{},~\text{other variations vanishing},\\
\bar{D}: & \delta \phi = -\bc \epsilon,~\delta c =(i \bphi -\frac{\beta}{\Gamma} \dot{\phi})\epsilon,~\delta \bphi=i\frac{\beta}{\Gamma} \dot{\bc} \epsilon  .~\text{other variations vanishing}
\end{array}
\label{eq:SuperDerivatives}
\end{equation}
If we write the theory in a manifestly supersymmetric way using superspace (see Appendix A), then these two operators play the role of (covariant) derivatives -- therefore we call them \emph{superderivatives}
To make connection with the symmetry algebra of the Schwinger-Keldysh construction for thermal initial states described in \cite{Haehl:2015foa} we note that upon identifications
\begin{equation}
\left\lbrace \begin{array}{l}
Q \equiv Q_{SK},\\
D \equiv \bar{Q}_{SK},\\
Q - \bar{D} \equiv Q_{KMS},  \\
D-\bar{Q} \equiv \bar{Q}_{KMS},\\
\end{array}\right. \label{eq:TransitionMukundZJ}
\end{equation}
we recover the charges connected with Schwinger-Keldysh formalism. The above were recently used to construct effective actions for dissipative hydrodynamics. A natural question arises whether they are true symmetries of the full partition function. We shall briefly present that, while $Q$ and $\bar{Q}$ are true symmetries, $D$ and $\bar{D}$ are symmetries of the partition function only for specific sources, which are explicitly invariant under these transformations. 

\subsection{Symmetries of the Langevin dynamics}
Symmetries of path integrals imply that various identities are satisfied by correlation functions. We will analyze the emergence of such identities in the Langevin dynamics. Before we do that let us recall the general procedure to derive these identities using the Schwinger-Dyson approach. We start with a general field theory defined by a path integral
\be \label{Eq:path}
Z[J]=\int [d\varphi] \exp \left[ -S(\varphi) +J\cdot \varphi \right].
\ee
Here $\varphi$ denotes the set of all fields in our action and $J\cdot \varphi$ is a source term.\footnote{The 'dot' product indicates that $J$ is a vector of sources -- one source per field -- and some sign subtleties and constants can be involved, like an overall $i$ factor traditionally in Quantum Field Theory or a minus sign for some Grassmanian source-field ordering.} We proceed by making an infinitesimal change of variables,
\be \label{Eq:inf}
\varphi(x)=\chi(x)+\epsilon F(x;\chi),
\ee
where $F(x;\chi)$ is a general functional of $\chi(x)$. It is enough to work to the linear order in $\epsilon$. The variation of the action functional takes the form
\be
S(\varphi)=S(\chi)+\epsilon \int \dx \frac{\delta S}{\delta \chi(x)}F(x;\chi) +O (\epsilon^2).
\ee
In addition to that, a general change of variables leads to a nonzero Jacobian,
\be
\mathcal{J}=\det \frac{\delta \varphi (x)}{\delta \chi (x)}=1+\epsilon \int \dx \frac{\delta F(x;\chi)}{\delta \chi(x)} +O (\epsilon^2).
\ee
The invariance of the path integral \eqref{Eq:path} under changes of variables means that the terms of order $\epsilon$ cancel out,
\begin{equation}
\int [d\varphi]\left(\int \dx \frac{\delta F(x;\chi)}{\delta \chi(x)}+\frac{\delta S}{\delta \chi(x)}F(x;\chi) + J\cdot F(x, \chi)  \right)\exp \left[ -S(\varphi) +\int\dd{x} J(x)\varphi(x) \right]  =0. \label{eq:SchwingerDysonNonDerForm}
\end{equation}
Now, we can use a field-theoretic trick: in a path integral, a field can be replaced by a variational derivative with respect to the corresponding source. For example, if we only had one field $\varphi$,
\begin{equation}
\int [d\varphi]\varphi(y) \exp \left[ -S(\varphi) +\int\dd{x} J(x)\varphi(x) \right] = \dv{J(y)}\int [d\varphi]\exp \left[ -S(\varphi) +\int\dd{x} J(x)\varphi(x) \right] .
\end{equation}
Combining the previous expressions and making the field-derivative replacement\footnote{$D_J$ denotes here the set of variational derivative operators corresponding to fields from $\chi$. Elements of $D_J$ are proportional to $\dv{J_i}$ (with $J_i$ being component of $J$) and proportionality constants take care of signs for Grassmans and the overall source term normalization.} $\chi \mapsto D_J$ we obtain the identity
\be\label{Eq:id}
\int \dx \left[ F(x;D_J) \frac{\delta S(D_J)}{\delta \chi (x)}- \frac{\delta F(x;D_J)}{\delta \chi (x)} -J(x) F(x;D_J) \right]Z[J]=0.
\ee
We see that if 
\be \label{Eq:JacandS}
\int \dx \left[ F(x;D_J) \frac{\delta S(D_J)}{\delta \chi (x)}- \frac{\delta F(x;D_J)}{\delta \chi (x)} \right]Z[J]=0,
\ee
then $F(x;D_J)$ generates a symmetry of the partition function. Expression \eqref{Eq:id} leads to Ward identities expressed in terms of currents,
\be
\int \dx \left[J(x) F(x;D_J) \right]Z[J]=0.
\ee
Ward identities can be satisfied if both terms in Eq. \eqref{Eq:JacandS} vanish separately or, in a more general case, if the term generated by the variation of the action is canceled by the term coming from the Jacobian. We note that it is a very common assumption that the symmetry of the action implies unit Jacobian. However, we will not make this assumption here. If $F(x;D_J)$ generates the symmetry of the action but it has a Jacobian that is not equal to one, then we say there is an anomaly,
\be
\int \dx \left[J(x) F(x;D_J) \right]Z[J]=\int \dx \left[- \frac{\delta F(x;D_J)}{\delta \chi (x)} \right]Z[J]\equiv \mathcal{A}.
\ee
We are now in a position to study the symmetries of the Langevin dynamics. To do that we will assume that the infinitesimal transformation takes the form
\be
\varphi =\varphi +\epsilon \mathcal{Q} \varphi, \qquad \mathcal{Q}\in \{Q,D,\bar{Q},\bar{D}\}
\ee
where $\varphi$ denotes the set of fields $(\phi, \bar{\phi}, c, \cb)$ and the generators $\mathcal{Q}$ are given by \eqref{eq:SymmetryAlgebra} and \eqref{eq:SuperDerivatives}.
We first investigate what is the transformation of the action \eqref{eq: MSRAction} for over-damped Langevin equation \eqref{eq:Langeven} those transformations (TTD -- total time derivative):
\begin{equation}
\left\lbrace \begin{array}{l}
\delta_Q \Sigma = 0,\\
\delta_D \Sigma = 2i c \dot{\bar{\phi}}=\delta_{\bar{Q}}(\bar{\phi}^2),\\
\delta_{\bar{Q}} \Sigma = 0+\text{TTD},\\
\delta_{\bar{D}} \Sigma =\left( \frac{2\beta}{\Gamma}\dot \phi-2i \bar{\phi}\right)\dot {\bar{c}}+\text{TTD} = \delta_{Q}(-2\frac{\beta}{\Gamma}\dot \phi^2-2 c \dot{\bar{c}})+\text{TTD}.\\
\end{array}\right. \label{eq:symmS}
\end{equation}
We see that $Q$ and $\bar{Q}$ generate symmetries of any time-independent action, while $D$ and $\bar{D}$ do not. However, the results of acting with $D$ and $\bar{D}$ on the action do not depend on the details of the theory -- they are independent of the potential. So, the identities generated by this change of variables should hold for every Langevin-type theory with a time-independent potential $U$. In case of an explicit time dependence in the equation, two of the above equations get modified. As noted in \cite{Orland}, the transformation $\delta_{\bar{Q}}$ ceases to be a symmetry, and in turn supersymmetry is broken out-of-equilibrium. Also, the last transformation law assumes the form
\begin{equation}
\delta_{\bar{D}} \Sigma =\left( \frac{2\beta}{\Gamma}\dot \phi-2i \bar{\phi}\right)\dot{\bar{c}} -\beta{}\bar{c}\frac{\p}{\partial t}\dv{\phi}U(\phi, t)+\text{TTD}, 
\end{equation}
and is no longer independent of the theory.\footnote{However, if the driving protocol involves linear coupling of field $\phi$ to some time dependent source $H(t, x)$, the discussion still holds as this coupling is technically identical to a source term in our effective action.} Another statement that we can make comes from the observation that the leftovers generated from the $D$ and $\bar{D}$ acting on the action functional can be obtained from symmetry transformations $\bar{Q}$ in the former case and $Q$ in the latter. As a result it leaves the full path integral invariant in the limit of vanishing sources, provided that the measure is invariant. This follows from an argument similar to the one used in the derivation of the supersymmetic localization technique \cite{Modave,Pestun:2016zxk}, which we present in Appendix \ref{sec:AppLoc}.  

Using the above reasoning in combination with \eqref{eq:SchwingerDysonNonDerForm}, we see that 
\begin{equation}
\int [d\varphi]\left(\int\dx~\delta_s H + J(x)\cdot\delta_d \varphi(x)   \right)\exp \left[ -S(\varphi) +\int\dd{x} J(x)\varphi(x) \right]=0 \label{eq:SDForSuperDeriv},
\end{equation}
where $\delta_s,~H$, and $\delta_d$ can be either :
$$
\delta_d=\delta_D,~\delta_s H=\delta_{\bar{Q}}(\bphi{}^2),
$$
or
$$
\delta_d=\delta_{\bar{D}},~\delta_s H=\delta_{Q}(-2\frac{\beta}{\Gamma}\dot{\phi}^2-2 c\cb).
$$
Now, one can observe that the first term of \eqref{eq:SDForSuperDeriv} can vanish as a full supersymmetric variation, if the following is satisfied:
\begin{equation}
\delta_s (J\cdot\varphi) = 0.
\end{equation}
This is of course not possible for a generic source term. However it can happen if one chooses a proper combination of sources or, alternatively, one might say that we choose to source an supersymmetric operator. In that case, we can replace the source term by
\begin{equation}
\int \dx~j(x)\mathcal{O(\varphi)},
\end{equation}
where now $j$ is a single source function and $\mathcal{O(\varphi)}$ is some function of fields that fulfils 
\begin{equation}
\delta_s\mathcal{O(\varphi)}=0.
\end{equation}
Since now both terms of \eqref{eq:SDForSuperDeriv} must vanish independently, we get some type of identity similar to Ward identities
\begin{equation}
\int [d\varphi]\left(\int\dx~j(x)\dv{\varphi(x)}\mathcal{O(\varphi)}\cdot{}\delta_d \varphi(x)\right)\exp \left[ -S(\varphi) +\int\dx~j(x)\mathcal{O(\varphi)} \right]=0.
\end{equation}
However, these identities are only valid for operators supersymmetric operators $\mathcal{O}$, so we chose to work further with more general expressions \eqref{eq:SDForSuperDeriv}
\section{Identities satisfied by Langevin dynamics}

We have seen how the supercharges and superderivatives transform the Langevin path integral. We now explore if they lead to new identities that emerge. In the case of supercharges the question is well understood. The path integral is invariant under the transformations $Q$ and $\bar{Q}$ which results in the corresponding Ward identities. Traditionally, Ward-Takahashi identities are identities between correlators of fields, derived from symmetry along with assumption that a path integral measure transforms with unit Jacobian, i.e. there is no quantum anomaly. They can be, however, stated in terms of identities between variational derivatives of the partition function, see for example \cite{Orland}. We are going to use those forms in our calculations.  The WT identities for symmetries \eqref{eq:SymmetryAlgebra} read
\begin{align}
&G_Q = \int \dx~\Hb(x)\dv{L(x)}{Z} +i \Lb(x)\dv{H(x)}Z=0, \\
&G_{\bar{Q}} = \int \dx~\Hb{(x)}\dv{\Lb(x)}Z -i\frac{\beta}{\Gamma}H(x) \partial_t \dv{\Lb(x)}Z - L\left[i \dv{H(x)}Z -\frac{\beta}{\Gamma} \partial_t \dv{\Hb(x)}Z\right]=0 ,
\label{eq:WardIdentities}
\end{align} 
where $x=(t, x_1, x_2, ...)$, and the number of spatial variables $x_i$ depends on the specific problem.

A natural question that emerges is whether the transformations coming from superderivatives also generate identities. Of course, for any transformation there exists a Schwinger-Dyson type of identity \eqref{Eq:id}, but we would like to see if there are simplifications or universalities that apply to this expression. One motivation for the existence of such a simplification comes from the analysis of \cite{Haehl:2015foa}. This work proposes to introduce a dynamical gauge field and to enlarge supersymmetric algebra by including also operators $D, \bar{D}$. This extension requires that the measure is invariant under $D$ and $\bar{D}$.
However, this is a subtle point and to justify it we will explicitly check it in a specific example of the Ornstein-Uhlenbeck process. Another observation that we can make follows from \eqref{eq:symmS}. The noninvariance under $D$ and $ \bar{D}$ comes purely from the kinetic term and leaves the potential term invariant. As a result for any choice of the (time-independent) potential the following two identities have to to be satisfied: 
\begin{align}
&\int \dx~-2i\dv{\Lb(x)}\p_t\dv{H(x)}Z=\int \dx~\left[-\Hb{(x)}\dv{\Lb(x)} +i \left(\dv{H(x)}\right) L(x)\right]Z \label{eq:NonWard1},\\
&\int \dx~2\left(\frac{\beta}{\Gamma}\p_t\dv{\Hb(x)} - i \dv{H(x)}\right)\p_t\dv{L(x)}Z =\int \dx \left[\Hb{(x)}\dv{L(x)} -i\frac{\beta}{\Gamma}H(x) \partial_t \dv{L(x)} + \Lb(x) \left(i \dv{H(x)} -\frac{\beta}{\Gamma} \partial_t \dv{\Hb(x)}\right)\right]{}Z\label{eq:NonWard2}.
\end{align}
Those two identities are different in character than standard Ward identities. The key difference is that there are no sources other than those in $Z$ on the left hand sides of (\ref{eq:NonWard1}, \ref{eq:NonWard2}). It means that no functional differentiation can remove the integrals from LHS. As a result the generated identities will have a sum rule form connecting local values of correlators from right-hand side of identities to integrals over the whole domain coming from the left-hand side (in Fourier space it translates to a integral over all frequencies). Interestingly, since $Z[J]$ is present both on the LHS and the RHS of identities, infinite amount of sum rules for higher-order correlators can be generated.
\subsection{Implications for correlators}
The identities (\ref{eq:NonWard1} and \ref{eq:NonWard2}) are written in a form that allows one to generate arbitrary amount of identities between correlators. It is, however, useful to see what kinds of identities for $n$-point functions can be obtained from them. First, we can just put all sources to zero, which causes the RHS of both identities to vanish [$\dot{\bphi}(x)=\p_t\bphi(x)$]:
\begin{align}
&\int{}\dx~2 i\cor{c(x)\dot{\bphi}(x)}=0,\\
&\int \dx~2\cor{\left(\frac{\beta}{\Gamma}\dot{\phi}(x) - i \bphi(x)\right)\dot{\cb}(x)}=0.
\end{align}
To see in a clearer way what kinds of sum rules can be obtained form the identities we take \eqref{eq:NonWard1} and apply operator $\dv{L(y)}$ to both sides and set the sources to zero to get
\begin{equation}
\int\dx~-2\left.\dv{L(y)}\dv{\Lb(x)}\p_t\dv{H(x)}Z\right\vert_{J=0} =\left. \dv{H(y)}Z\right\vert_{J=0}. \label{eq:SumStep1}
\end{equation} 
Now, to get rid of $\dv{L(y)}\dv{\Lb(x)}$ we can use $G_Q$ of \eqref{eq:WardIdentities},
\begin{equation}
\left.\dv{\Lb(x)}\dv{\Hb(y)}\dv{H(z)}G_Q\right\vert_{\Lb=0,\Hb=0}=~\left.\dv{H(z)}\left(\dv{\Lb(x)}\dv{L(y)}+i\dv{\Hb(y)}\dv{H(x)}\right)Z\right\vert_{\Lb=0,\Hb=0}=0.
\end{equation}
Plugging the above into \eqref{eq:SumStep1} gives
\begin{equation}
2i \int\dx~\left.\dv{H(x)}\dv{\Hb(y)}\p{}_t\dv{H(x)}Z\right\vert_{L,\Lb,\Hb=0} = \left.\dv{H(y)}Z\right\vert_{L,\Lb,\Hb=0},
\label{eq:CorrRelation2}
\end{equation}
which in terms of correlation functions is
\begin{equation}
2i\int\dx~\cor{\bphi(x)\phi(y)\dot{\bphi}(x)} = \cor{\bphi(y)}.
\end{equation}
Since $\cor{\bphi(y)}=0$ (as a consequence of $G_Q=0$), we see, that integral on LHS of the above vanishes. If we introduce Fourier transforms of fields
\begin{equation}
\phi(k) = \int\dx e^{-ik\cdot{}x}\phi(k),
\end{equation}
we get a sum-rule-type relation,
\begin{equation}
\forall_q~\int\dd{k}~\omega\cor{\bphi(-k)\phi(q)\bphi(k)}=0,
\end{equation}
where $\omega$ denotes first component of $k$ -- the conjugate of $t$.
Alternatively, relation \eqref{eq:CorrRelation2} can be rewritten as a response function identity
\begin{equation}\label{eq:nlresponse}
\int\dx~\dv{H(x)}\der{t}\dv{H(x)}\cor{\phi(y)} = 0.
\end{equation}
The Ward identities \eqref{eq:WardIdentities} lead to the equilibrium relations between observables. New identities \eqref{eq:NonWard1} and \eqref{eq:NonWard2} provide additional constraints among correlation functions. These constraints should be placed among results following from generalizations of linear responses to higher orders \cite{2007Gaspard,2015Basu}. Usually nonlinear responses depend on the details of the dynamics. Therefore, universal nonlinear relations are rather remarkable. \emph{A priori} it is not evident that such a relation should exist. We see that the relation \eqref{eq:nlresponse} does not relate correlations to responses as is the case with the fluctuation-dissipation theorem. Instead it places a constraint that integrated non-linear response vanishes. Furthermore by acting with more derivatives we can generate universal relations valid for higher-order response functions. The consequences of this constraint are beyond the supersymmetric formalism used here. Nevertheless, we can show that the relation holds in an explicit example of the Gaussian potential.

\subsection{Gaussian theory: Ornstein-Uhlenbeck process}
The Ornstein-Uhlenbeck process, describing amongst others thermal noise in RLC circuits, is the simplest of Langevin dynamics and under analytic control. It is defined as
\begin{equation}
E[\phi(t)] =\dot{\phi(t)} + \Gamma \phi(t) = \nu(t) \label{eq:OUProcess}
\end{equation}
i.e. it is one-dimensional Gaussian model, in a sense that the fields appear in the effective action at most in second powers. We can calculate the partition function by doing the Gaussian integrals in bosonic and fermionic fields separately
\be
Z[J]=Z[\Hb, H, \Lb, L]=Z_b[\Hb, H]Z_f[\Lb, L] .
\ee
The result is given by
\begin{align}
Z_b[\Hb, H]&= \exp\left( \int \dd\tau\dd\tau{}^\prime e^{-\Gamma|\tau - \tau{}^\prime|}\frac{1}{2\beta}\Hb(\tau)\Hb(\tau{}') - i\theta(\tau - \tau^\prime) e^{-\Gamma (\tau - \tau^\prime)}\Hb(\tau{})H(\tau{}')\right), \\
Z_f[\Lb, L]  &= \exp\left( -\int \dd\tau\dd\tau{}^\prime L(\tau)\theta(\tau- \tau^\prime)e^{-\Gamma(\tau - \tau{}^\prime)}\Lb(\tau{}')\right).\label{eq:PartFuncUO}
\end{align}
with $\theta$ being a Heaviside step function and $i$ -- imaginary unit. \\
Upon using the partition function \eqref{eq:PartFuncUO} one can evaluate all identities directly and find them to be satisfied -- see Appendix \ref{sec:AppBCheck}.

\section{Summary and outlook}
In this paper we studied the transformations of $\mathcal{N}_T=2$ algebra acting on the Langevin dynamics formulated in terms of a supersymmetric path integral. We found that two operators $D$ and $\bar{D}$, despite not being symmetries of the action, generate additional and universal identities since they change action in a potential-independent way. In the previous studies identities among the correlation functions were identified to be equivalent with the equilibrium relations among the correlation functions. In addition to that, such identities have consequences in the nonequilibrium dynamics. In this case, the supersymmetry is violated; however, it can be partially recovered by adding to the dynamical action a term which corresponds to Jarzynski's work \cite{Jarzynski1997}. A natural extension of this work is to check how time-dependent potentials modify the new relations coming from transformations due to superderivatives.

We have shown that if we eliminate ghost fields, then the relations we obtain result in a nonlinear and nonlocal constraint on response functions. This is an important difference with respect to the relations coming from supersymmetric Ward identities. It can be traced to the fact that a general nonlinear response depends on the underpinning dynamics. However, the relations that we obtain capture only the universal correlation functions, independent of the dynamics. To explore physical consequences of such relations one needs to go beyond the supersymmetric formalism employed in this note. 

Another direction that can be studied in more detail is the Langevin dynamics with colorful noise. It was shown \cite{Aron} that the structure that emerges in this case resembles closely Langevin dynamics with the noise and the identities following from supersymmetry hold in these generalised case. However, we stress that the inclusion of time dependence and colorful noise is not automatic. It may happen that the regularization procedure or the properties of the fermionic determinant imply a breaking of identities \ref{eq:NonWard1}.

We have demonstrated that the supersymmetric identities hold in Ornstein-Uhlenbeck process. These systems has been studied with different theoretical approaches and it is also easily accessible experimentally. Therefore, it can serve as a playground to get more intuition about the relations we derived.

Finally, the same structure is present in constructions of the effective actions for fluids. The simplest fluid is described by Burgers equation, which is a Langevin dynamics that is not potential. A detailed analysis of the identities derived here could shed new light on various correlates in fluid dynamics.

\section*{Acknowledgements} 
We acknowledge useful discussions with Andre Cardoso Barato, Charlie Duclut, Guido Festuccia, Mukund Rangamani, and Udo Seifert. We are especially grateful to Kristan Jensen for pointing out errors in the previous version of the manuscript. We would also like to thank the referees of this paper, whose critical remarks allowed us to improve the structure and presentation. This work was supported by the Deutsche Forschungsgemeinschaft via the Leibniz Programm.
\appendix
\section{Superspace formulation}
Since in the text we make a few references to superspace objects (superderivatives, superspace translations), we briefly show the superspace formulation of our theory of interest. The superspace is a space spanned by physical dimesions and some number of grassmann directions which are basis elements of a skew-symmetric grassman algebra -- in our case $\theta, \bar{\theta}$.
 We begin with the superfield -- object that encodes our fields:
\begin{equation}
\Phi = \phi+ \theta\cb+c\btheta + \theta\btheta i \bphi.
\end{equation}
To have the proper source term namely
\begin{equation}
\int\dx\dd{\btheta}\dd{\theta}~J \Phi = \int\dx~\Hb\phi + H\bphi + L\cb + c\Lb,
\end{equation}
 our super source $J$ must be
\begin{equation}
J=-i H + \Lb\theta+\btheta L + \theta\btheta \Hb.
\end{equation}
The symmetry generators are listed in the table below:
\begin{equation}
\begin{array}{l l}
D=\partial_{\btheta}, & \bar{D} = \partial_{\theta} -\frac{\beta}{\Gamma}\btheta \der{t}, \\
~&~\\
Q=\partial_\theta, & \bar{Q} = \partial_{\btheta} + \frac{\beta}{\Gamma}\theta \der{t} ,\\
~&~\\
\bar{Q} = \p_{\bar{\theta}}+\frac{\beta}{\Gamma}\theta \der{t}, & \bar{D}=\p_{\theta}-\frac{\beta}{\Gamma}\btheta \der{t},
\end{array}
\end{equation}
Now partition function \eqref{eq:PartFu} for a system of our interest can be written as
\begin{equation}
Z[H, \Hb, L, \Lb]=\int \D{\Phi} \exp\left(\int\dx\dd{\btheta}\dd{\theta}~-\Gamma\left(\frac{1}{\beta}\bar{D}\Phi{}D\Phi + U(\Phi)\right)+J\Phi\right).\label{eq:SUSYPart}
\end{equation}
Now it is clear  why we call $D,~\bar{D}$ "superderivatives" -- they appear in the superspace kinetic term.

\section{Supersymmetric localization}\label{sec:AppLoc}
The idea behind supersymmetric localization resembles a saddle-point approximation of integrals, however, it yields exact results provided that the partition function is invariant under a supersymmetry. Let us assume that we have a theory defined by a path-integral over both bosonic ($\phi$) and fermionic ($\psi$) fields
\begin{equation}
Z=\int \D{\phi}\D{\psi} \exp{(S[\phi,\psi])} \label{eq:ActionGeneral}
\end{equation}
where $S$ is the supersymmetric action and the path integral measure transforms under change of variables given by $\delta_s$ with a unit Jacobian. Also, let $\delta_s$ be a fermionic symmetry transformation and $\delta_s^2 = B$ -- a bosonic transformation and $V[\phi,\psi]$ a functional of fields, such that $BV=\delta_s^2 V =0$ and bosonic part of $\delta{}V$ is positive.  Then the supersymmetric localization principle states that deforming our action with a term proportional to $V$ does not change partition function
\begin{equation}
\der{\mu} \int \D{\phi}\D{\psi} \exp{\left(S[\phi,\psi]+ \mu \delta_s V[\phi,\psi]\right)} = 0.\label{eq:locTheorem}
\end{equation}
This is a consequence of the relation 
\begin{equation}
\der{\mu}\int \D{\phi}\D{\psi} \exp{\left(S[\phi,\psi]+ \mu \delta_s V[\phi,\psi]\right)}=-\int \D{\phi}\D{\psi} \delta_s\left(V[\phi,\psi]\exp{\left(S[\phi,\psi]+ \mu \delta_s V[\phi,\psi]\right)}\right)=0,
\end{equation}
where the last equality follows from the fact, that functional integral of full supersymmetric variation vanishes \cite[see eq. 3.4]{Modave}
Equation \eqref{eq:locTheorem} shows also that a deformation of action proportional to a symmetry charge does not contribute to the path integral. Let us explore the localization technique further. If we let $\mu\rightarrow \infty$, $Z$ reduces to an integral over critical points of $\delta_s V$ and a small fluctuation (of WKB type) around them. The set of critical points of $\delta_s V$ is called localization locus and the part coming from small fluctuations -- 1-loop determinant. For many supersymmetric theories, especially on compact manifolds, the localization principle allows us to reduce the path-integral to a finite dimensional integral. For practical reasons one usually uses $V$  in the form
\begin{equation}
V = \sum\limits_{i} \psi_i\cdot{}(\delta_s \psi_i),\label{eq:standardV}
\end{equation}
where the sum runs over all fermions in the theory and $"\cdot{}"$ denotes some scalar product (depending on a theory, it may require for example the use of Hermitian conjugation). Then 
\begin{equation}
\delta_s V\vert_{\text{bos}} = \sum\limits_{i} (\delta_s \psi_i)\cdot{}(\delta_s \psi_i),
\end{equation}
which is manifestly positive and quadratic so that the locus is given by
\begin{equation}
(\delta_s \psi_i)\vert_{\psi=\psi_0,~\phi=\phi_0} = 0.
\end{equation}
Here, we denote locus field configuration as $\psi_0,~\phi_0$. These fields will dominate the path integral in the limit $\mu\rightarrow\infty$. To compute potential corrections, co-called 1-loop corrections, we redefine fields
\begin{equation}
\psi_i= \psi_{0,i} + \mu^{-1/2} \tilde{\psi}_i
\end{equation}
and then what survives the large $t$ limit is
\begin{multline}
Z=\int\limits_{\text{loc}} \D{\phi_0}\D{\psi_0} \exp{(S[\phi_0,\psi_0])}\int\D{\tilde{\phi}}\D{\tilde{\psi}} \exp{\left(\frac{\delta{}^2}{\delta\phi\delta\phi}\delta_s V\vert_{\phi_0}\right)}=\\
\int\limits_{\text{loc}} \D{\phi_0}\D{\psi_0} \exp{(S[\phi_0,\psi_0])}\text{SDet}(\delta_sV[\phi_0)]^{-2},\label{eq:LocRes}
\end{multline}
where the last line is just changing notation to more compact one. Let us also note, that this method allows us to compute correlation functions of supersymmetric operators, i.e ones for which
\begin{equation}
\delta_s \mathcal{O} = 0,\label{eq:SusyOperatorCondition}
\end{equation}
where $\mathcal{O}$ is the operator. This can be easily seen by observing that deforming action in \eqref{eq:ActionGeneral} by a source term for such an operator,
\begin{equation}
S[\phi,\psi] \rightarrow\tilde{S}[\phi,\psi, J]=S[\phi,\psi]+\int\dd{x} J \mathcal{O}
\end{equation}
does not affect any of the assumptions mentioned before \eqref{eq:locTheorem}. 
One could in principle use this technique to compute values of some operators composed of fields $(\phi, ~\bphi, ~c, ~\cb)$ of \eqref{eq: MSRAction}. However, condition \eqref{eq:SusyOperatorCondition}, where $\delta_s=\alpha \delta_{Q}+\beta \delta_{\bar{Q}}$ is some linear combination of symmetry generators \eqref{eq:SymmetryAlgebra} limits practical applications of this method. 

\section{Evaluation of an identity for the Ornstein-Uhlenbeck process}\label{sec:AppBCheck}

We shall sketch necessary steps in the computation of supersymmetric identities in the Ornstein-Uhlenbeck process on the example of \eqref{eq:NonWard1}. From \eqref{eq:PartFuncUO} we see:
\begin{align}
& \dv{H(t)}Z = \int\dd{t_1}~-i\theta(t_1-t) e^{-\Gamma (t_1-t)}\Hb{(t_1)} Z,\\
& \dv{\Lb(t)} = \int\dd{t_2}~\theta(t_2-t) e^{-\Gamma (t_2-t)}\Hb{(t_2)} Z.
\end{align}
With that, we can write the LHS of \eqref{eq:NonWard1} as
\begin{multline}
2\int \dd{t_1}\dd{t_2}\dd{t}~ L(t_2)\Hb(t_1)\theta{(t_2-t)}e^{-\Gamma (t_2-t)}*\der{t}\left[\theta(t_1-t) e^{-\Gamma (t_1-t)}\Hb{(t_1)} \right] \\=\int \dd{t_1}\dd{t_2}~L(t_2)\Hb(t_1)\left[e^{-\Gamma|t_1-t_2|}-2\theta(t_2-t_1) e^{-\Gamma (t_2-t_1)}\right].\label{eq:IdLHS}
\end{multline}
In evaluating the above we took into account that $\der{t}\theta{(t)}=\delta(t)$ and evaluated $t-$integrals in both terms coming from the time derivative. Next, we compute the RHS of the identity \eqref{eq:NonWard1} as
\begin{equation}
\int \dd{t_1}\dd{t_2}~e^{-\Gamma(t_1-t_2)}\theta{(t_1-t_2)}\left[L(t_1)\Hb(t_2)-L(t_2)\Hb(t_1) \right].\label{eq:IdRHS}
\end{equation}
Now, one can observe that the term with $\theta$ in \eqref{eq:IdLHS} can be subtracted from one of \eqref{eq:IdRHS} terms, to yield a symmetric combination. That means the the identity can be put in the following form
\begin{equation}
\int\dd{t_1}\dd{t_2}~L(t_2)\Hb(t_1)e^{-\Gamma|t_1-t_2|}=\int \dd{t_1}\dd{t_2}~e^{-\Gamma(t_1-t_2)}\theta{(t_1-t_2)}\left[L(t_1)\Hb(t_2)+L(t_2)\Hb(t_1) \right].
\end{equation}
To see that the last equation indeed holds for every functions $L,~\Hb$ is rather straightforward -- one must change integration variables in first term in ($t_1\leftrightarrow{}t_2$) and see, that it results in integrating sources with
\begin{equation}
\left\lbrace\begin{array}{c}
e^{-\Gamma (t_1-t_2)}~\text{if}~t_1>t_2,\\
e^{-\Gamma (t_2-t_1)}~\text{if}~t_2>t_1,
\end{array} \right.
\end{equation}
which is explicitly the same as on the RHS,\footnote{Set $t_1=t_2$ is zero measure and does not affect the integral}, which was to be shown. Using similar transformations, one can show that the identity \eqref{eq:NonWard2} holds as well.

\bibliographystyle{apsrev4-1}
\bibliography{langevin-bib}

\end{document}